\documentclass[9pt,conference]{IEEEtran}

\usepackage{waspaa25} 

\usepackage{bm} 
\usepackage{mathtools}

\newcommand{\Add}[1]{#1}

\usepackage{fancyhdr}

\fancypagestyle{firstpage}{%
  \fancyhf{}                     

  \fancyfoot[C]{%
    \fbox{
      \begin{minipage}{2\columnwidth}
        \footnotesize
        This paper has been accepted to the IEEE Workshop on Applications of Signal Processing to Audio and Acoustics (WASPAA) 2025.\\
        © 2025 IEEE. Personal use of this material is permitted. Permission from IEEE must be obtained for all other uses, in any current or future media, including reprinting/republishing this material for advertising or promotional purposes, creating new collective works, for resale or redistribution to servers or lists, or reuse of any copyrighted component of this work in other works.
      \end{minipage}
    }%
  }
}
\title{Kernel ridge regression based sound field estimation using a rigid spherical microphone array}

\name{Ryo Matsuda$^{1}$,
      Juliano G. C. Ribeiro$^{1}$,
      Hitoshi Akiyama$^{1}$,
      Jorge Trevino$^{1}$}
\address{$^{1}$Yamaha Corporation, Hamamatsu, Japan}

\begin{document}

\maketitle
\thispagestyle{firstpage}
\begin{abstract}
We propose a sound field estimation method based on kernel ridge regression using a rigid spherical microphone array. Kernel ridge regression with physically constrained kernel functions, and further with kernel functions adapted to observed sound fields, have proven to be powerful tools. However, such methods generally assume an open-sphere microphone array configuration, i.e., no scatterers exist within the observation or estimation region. Alternatively, some approaches assume the presence of scatterers and attempt to eliminate their influence through a least-squares formulation. Even then, these methods typically do not incorporate the boundary conditions of the scatterers, which are not presumed to be known.

In contrast, we exploit the fact the scatterer here is a rigid sphere. Meaning, both the virtual scattering source locations and the boundary conditions are well-defined. Based on this, we formulate the scattered sound field within the kernel ridge regression framework and propose a novel sound field representation incorporating a boundary constraint. The effectiveness of the proposed method is demonstrated through numerical simulations and real-world experiments using a newly developed spherical microphone array.
\end{abstract}

\section{Introduction}
\label{sec:1}
Estimating the spatial distribution of sound pressure, i.e., the sound field, from observations obtained using multiple microphones or microphone arrays is an essential technology for many applications in acoustics, such as spatial audio reproduction \cite{943347,poletti2005three,8798873,wu2010spatial,salvador2017boundary,otani2020binaural,matsuda2021binaural}, room \Add{acoustics} analysis \cite{nolan2018wavenumber, tanaka2023isotropic}. It is well known that the sound field in the frequency domain follows the Helmholtz equation and can be represented as a superposition of basis functions such as spherical wave functions (SWFs) \cite{eWilliams1999,poletti2005three} or plane waves \cite{rafaely2004plane}, which are general solutions to the equation. Traditionally, sound field estimation has been formulated as a linear regression problem in which the coefficients of these basis function expansions are estimated \cite{park2005sound,poletti2005three,abhayapala2002theory}.

With the recent advancements in deep learning, methods have been proposed to estimate sound fields by learning basis functions or regression processes from large-scale observational data \cite{lluis2020sound,miotello2024reconstruction}. However, these methods are not only constrained by the requirement that microphone positions must lie on a grid, but it is also not always feasible to prepare an appropriate dataset in advance. Meanwhile, kernel ridge regression (KRR), which estimates the sound field as a regression problem defined in a reproducing kernel Hilbert space, does not have such restrictions. These methods have shown that kernel functions incorporating the physical constraints enhance the estimation accuracy compared to kernels without physical constraints \cite{ueno2017sound,8521334,9394763,caviedes2021gaussian}. Furthermore, \Add{estimation performance can be improved by optimizing the kernel parameters using prior or estimated directional information of incoming waveftonts} \cite{10693558,ribeiro2024physics}.

However, the existing KRR methods cannot be directly applied to sound field estimation in the presence of scatterers such as a rigid spherical microphone array (RSMA)\cite{kaneko2018development}, even though use of RSMAs has been increasing\cite{rafaely2004analysis}. While previous studies have proposed interpolation methods that account for scattering objects\cite{koyama2023kernel,kozuka2024sound}, no method has been introduced to explicitly incorporate boundary conditions when they are well-defined. \Add{On the other hand, physics-informed neural network-based estimation methods for RSMA have been proposed \cite{10317164,10694489}. However, it is difficult to incorporate prior information in these methods, such as source direction, making them more susceptible to spatial aliasing.}

In this study, we propose a sound field estimation method that enables the use and optimization of the conventional kernel functions within the framework of KRR by introducing the kernel representation of the scattered sound field and a boundary constraint on a RSMA. We demonstrate the effectiveness of the proposed method by applying it to numerical simulations and real-world experiments in an anechoic chamber using our newly developed RSMA.

\section{Problem setting}
\label{sec:2}
Let us define a region of interest in three-dimensional space $\Omega \subset \mathbb{R}^3$, and denote the sound pressure as $u(\mathbf r,k)$, where $\mathbf r\in\mathbb R^3$ is the position, and $k=2\pi f$ is the wave number associated with the frequency $f$ and sound speed $c$.
Assuming that the sound field $u(\mathbf r,k)$ in the target region $\Omega$ consists of the incident sound field $u_{\rm inc}(\mathbf r,k)$ and the sound field $u_{\rm sct}(\mathbf r,k)$ generated by a scatterer (i.e., a rigid microphone array) in $\Omega_{\rm sct}\subset\Omega$, the total sound field is modeled as
\begin{equation}
u(\mathbf r,k) = u_{\rm inc}(\mathbf r,k)+u_{\rm sct}(\mathbf r,k).
\label{eq:1}
\end{equation}

Our objective is to estimate $u_{\rm inc}(\mathbf r,k)$ from the sound pressures $\{s_m\}_{m=1}^M$ measured by $M$ microphone units at $\{\mathbf r_m\}_{m=1}^M$ on a RSMA with a radius of $R$.

\section{Conventional methods}
\label{sec:3}
\subsection{SWF expansion with boundary conditions}
\label{sec:3:1}
The incident and scattered sound fields in (\ref{eq:1}) can be expressed in the SWF domain as \cite{eWilliams1999}
\begin{equation}
\begin{aligned}
u_{\rm inc}(\mathbf r) &= \sum_{\nu=0}^\infty\sum_{\mu=-\nu}^\nu \mathring{u}_{\rm inc,\nu,\mu}\underbrace{j_\nu(k\|\mathbf r\|)Y_{\nu,\mu}(\mathbf r/\|\mathbf r\|)}_{\varphi_{\nu,\mu}(\mathbf r)},\\
u_{\rm sct}(\mathbf r) &= \sum_{\nu=0}^\infty\sum_{\mu=-\nu}^\nu \mathring{u}_{\rm sct,\nu,\mu}\underbrace{h_\nu(k\|\mathbf r\|)Y_{\nu,\mu}(\mathbf r/\|\mathbf r\|)}_{\psi_{\nu,\mu}(\mathbf r)},\\
\end{aligned}
\label{eq:2}
\end{equation}
where $j_\nu(\cdot)$ is the $\nu$th-order spherical Bessel function, $h_\nu(\cdot)$ is the $\nu$th-order spherical Hankel function of the first kind, and $Y_{\nu,\mu}(\cdot)$ is the spherical harmonic function of order $\nu$ and degree $\mu$, respectively. The infinite series in (\ref{eq:2}) is truncated at some order $N$ hereafter. Using the Neumann boundary conditions \cite{eWilliams1999} on $\Omega_{\rm sct}$ expressed as
\begin{equation}
\frac{\partial u(\mathbf r)}{\partial \mathbf n(\mathbf r)}\Bigg|_{r=R}=0
\label{eq:3}
\end{equation}
where $\mathbf n(\mathbf r)$ denotes the normal direction of $\Omega_{\rm sct}$ at $\mathbf r$, the sound pressure at the microphone on $\partial\Omega_{\rm sct}$ can be represented as \cite{poletti2005three}
\begin{equation}
u(\mathbf r_m)\Add{\simeq}\sum_{\nu=0}^N\sum_{\mu=-\nu}^\nu \mathring{u}_{\rm inc,\nu,\mu}\underbrace{\frac{\rm j}{(kR)^2h'_\nu(kR)}Y_{\nu,\mu}(\mathbf r_m/\|\mathbf r_m\|)}_{C_{\nu,\mu}(\mathbf r_m)}.
\label{eq:4}
\end{equation}
Then, the objective function is expressed as a minimization problem with a reguralization as
\begin{equation}
\underset{\mathring{\mathbf u}_{\rm inc}}{\rm minimize}\ \mathcal J(\mathring{\mathbf u}_{\rm inc})\coloneqq\|\mathbf s-\mathbf C\mathring{\mathbf u}_{\rm inc}\|^2+\lambda\mathring{\mathbf u}_{\rm inc}^\mathsf H\mathring{\mathbf u}_{\rm inc},
\label{eq:5}
\end{equation}
where $\mathbf s=[s_1,\dots,s_M]^\mathsf T\in\mathbb C^M$, $\mathring{\mathbf u}_{\rm inc}=[\mathring{u}_{{\rm inc},0,0},\dots,\mathring{u}_{{\rm inc},N,N}]^\mathsf T\in\mathbb C^{(N+1)^2}$, $\mathbf C\in\mathbb C^{M\times(N+1)^2}$ is the matrix whose elements are $[\mathbf C]_{m,\nu^2+\nu+\mu+1}=C_{\nu,\mu}(\mathbf r_m)$, and $\lambda\in\mathbb R_{\geq0}$ is the regularization parameter. The solution of (\ref{eq:5}) can be obtained as
\begin{equation}
\hat{\mathring{\mathbf u}}_{\rm inc}=(\mathbf C^\mathsf H\mathbf C+\lambda \mathbf I)^{-1}\mathbf C^\mathsf H\mathbf s.
\label{eq:6}
\end{equation}
and then the incident sound field is obtained by using (\ref{eq:2}) and (\ref{eq:6}).

\subsection{KRR without boundary conditions}
\label{sec:3:2}
Unlike the SWF expansion, which requires determining the maximum order $N$ on an empirical basis, the incident sound field can also be represented in a form of KRR as a result of the infinite series expansion \cite{ueno2017sound,8521334} as
\begin{equation}
u_{\rm inc}(\mathbf r,k)\Add{\simeq}\sum_{m=1}^M\alpha_m\kappa_{\rm inc}(\mathbf r,\mathbf r_m),
\label{eq:7}
\end{equation}
where $\alpha_m\in\mathbb C$ denotes the coefficients of KRR and $\kappa_{\rm inc}(\mathbf r,\mathbf r')$ denotes the kernel function. As a physically constrained kernel function, the Bessel kernel \cite{ueno2017sound,8521334} can be used when assuming a diffuse sound field, given by 
\begin{equation}
\kappa^{\rm Bs}_{\rm inc}(\mathbf r,\mathbf r')=j_0(k\|\mathbf r-\mathbf r'\|).
\label{eq:8}
\end{equation}
When prior information or optimization allows the utilization of the direction of arrival of the wavefront, the multi directional (MD) kernel \cite{9394763,10693558,ribeiro2024physics} can be employed, given by 
\begin{equation}
\begin{aligned}
&\kappa^{\rm MD}_{\rm inc}(\mathbf r,\mathbf r')\\
&=\sum_{q=1}^Q\gamma_q\frac{j_0(\sqrt{(k(\mathbf r-\mathbf r')-{\rm j}\zeta_q\mathbf d_q)^\mathsf T(k(\mathbf r-\mathbf r')-{\rm j}\zeta_q\mathbf d_q)})}{C(\zeta_q)},
\end{aligned}
\label{eq:9}
\end{equation}
where 
\begin{equation}
C(\zeta)= \begin{cases}1, & \zeta=0 \\ \frac{\sinh (\zeta)}{\zeta}, & \text { otherwise }\end{cases},
\label{eq:10}
\end{equation}
$Q\in\mathbb N$ is the number of directions, $\mathbf d_q\in\mathbb R^3$ is the $q$th bias direction, $\zeta_{q}\in\mathbb R_{\geq0}$ is the concentration parameter for the $q$th bias direction.
Then, the total sound field can be expressed by sum of (\ref{eq:7}) and $u_{\rm sct}$ in (\ref{eq:2}), and the minimization problem is formulated as \cite{koyama2023kernel}
\begin{equation}
\begin{aligned}
\underset{\boldsymbol \alpha, \mathring{\mathbf u}_{\rm sct}}{\rm minimize}\ \mathcal J(\boldsymbol \alpha, \mathring{\mathbf u}_{\rm sct})&\\
\coloneqq\|\mathbf s-\mathbf K_{\rm inc}\boldsymbol \alpha-\mathbf \Psi&\mathring{\mathbf u}_{\rm sct}\|^2+\lambda_1\boldsymbol \alpha^\mathsf H\mathbf K_{\rm inc}\boldsymbol \alpha+\lambda_2\mathring{\mathbf u}_{\rm sct}^\mathsf H\mathbf W\mathring{\mathbf u}_{\rm sct},
\end{aligned}
\label{eq:11}
\end{equation}
where $\boldsymbol\alpha=[\alpha_1,\dots,\alpha_M]^\mathsf T\in\mathbb C^M$, $\mathring{\mathbf u}_{\rm sct}=[\mathring{u}_{{\rm sct},0,0},\dots,\mathring{u}_{{\rm sct},N,N}]^\mathsf T\in\mathbb C^{(N+1)^2}$, $\mathbf K_{\rm inc}\in\mathbb C^{M\times M}$ is the Gram matrix whose elements are $[\mathbf K_{\rm inc}]_{m,m'}=\kappa_{\rm inc}(\mathbf r_m,\mathbf r_{m'})$, and $\mathbf \Psi\in\mathbb C^{M\times (N+1)^2}$ is the basis matrix whose elements are $[\mathbf \Psi]_{m,\nu^2+\nu+\mu+1}=\psi_{\nu,\mu}(\mathbf r_m)$. $\mathbf W\in\mathbb C^{(N+1)^2\times(N+1)^2}$ is the weighting matrix assuming the smoothness of the scattered field \cite{koyama2023kernel}, and $\lambda_1\in\mathbb R_{\geq 0}$, $\lambda_2\in\mathbb R_{\geq 0}$ are the regularization parameters.
We can solve (\ref{eq:11}) for ${\boldsymbol{\alpha}}$ as \cite{koyama2023kernel,kozuka2024sound}
\begin{equation}
\hat{\boldsymbol{\alpha}}=\left(\mathbf{K}_{\rm inc}+\lambda_1 \mathbf{I}+\frac{\lambda_1}{\lambda_2} \boldsymbol{\Psi} \mathbf{W}^{-1} \boldsymbol{\Psi}^{\mathsf{H}}\right)^{-1} \mathbf{s},
\label{eq:12}
\end{equation}
and the incident sound field is estimated using (\ref{eq:7}). 

\section{Proposed method}
\label{sec:format}
\subsection{Model}
\label{sec:4:1}
To represent the scattered sound field, we introduce the source region (SR) kernel \cite{7327167,matsuda2025sound}, which is the spatial correlation of all point sources within a scatterer, expressed as
\begin{equation}
\begin{aligned}
&\kappa_{\rm sct}(\mathbf r,\mathbf r')=\int_{\mathbf r_s\in\Omega_{\rm sct}}G(\mathbf r|\mathbf r_s)\overline{G(\mathbf r'|\mathbf r_s)}w(r_s){\rm d}\mathbf r_s \\
&=\sum_{\nu=0}^\infty\sum_{\mu=-\nu}^\nu \psi_{\nu,\mu}(\mathbf r)\overline{\psi_{\nu,\mu}(\mathbf r')}\underbrace{k^2\int_0^Rr_s^2j_\nu(kr_s)^2w(r_s){\rm d}r_s}_{\xi_\nu(R)}
\end{aligned}
\label{eq:13}
\end{equation}
where $G(\mathbf r|\mathbf r_s)$ denotes the three-dimensional free-field Green's function, $w(r_s)$ denotes the spherically symmetric probability distribution of sound sources inside $\Omega_{\rm sct}$. The infinite series in (\ref{eq:13}) is truncated at $N_{\rm ext}$, hereafter. Assuming $w(r_s)$ as a uniform source distribution, a weight function $\xi_\nu(R)$, obtained as a result of integration in the second line of (\ref{eq:13}), can be derived analytically as \cite{8798873,7327167,matsuda2025sound}
\begin{equation}
\xi_\nu(R)=\frac{3k^2}{8\pi}\left[j_\nu\left(k R\right)^2-j_{\nu-1}\left(k R\right) j_{\nu+1}\left(k R\right)\right].
\label{eq:14}
\end{equation}
Here, the representation of the scattered field using the SR kernel becomes equivalent to the one using the SWF if we set $\xi_\nu(R)=1$ \cite{matsuda2025sound}. Then, the scattered field can be represented in a KRR form as
\begin{equation}
u_{\rm sct}(\mathbf r,k)\Add{\simeq}\sum_{m=1}^M\beta_m\kappa_{\rm sct}(\mathbf r,\mathbf r_m).
\label{eq:15}
\end{equation}
and the total sound field can be modeled as 
\begin{equation}
\mathbf s=\mathbf K_{\rm inc}\boldsymbol\alpha+\mathbf K_{\rm sct}\boldsymbol\beta+\boldsymbol\varepsilon,
\label{eq:16}
\end{equation}
where $\mathbf K_{\rm sct}\in\mathbb C^{M\times M}$ is the Gram matrix whose elements are $[\mathbf K_{\rm sct}]_{m,m'}=\kappa_{\rm sct}(\mathbf r_m,\mathbf r_{m'})$, $\boldsymbol{\beta}=[\beta_1,\dots,\beta_M]\in\mathbb C^M$ is the vector of coefficients, and $\boldsymbol\varepsilon\in\mathbb C^M$ is the measurement noise. \Add{The difference between the SWF expansion and the proposed model lies in the type of constraint imposed: the SWF expansion enforces a truncation in the expansion order, whereas the proposed model constrains the feature space. Moreover, the SWFs serve as finite-order isotropic estimators, while the proposed kernels are infinite-order anisotropic estimators that can incorporate source direction biasing, as in Eq.~(\ref{eq:9}).}

\subsection{A Boundary constraint and the optimization problem}
\label{sec:4:2}
In order to incorporate the Neumann boundary conditions on $\partial \Omega_{\rm sct}$ into KRR, the boundary condition of (\ref{eq:3}) is replaced with a soft constraint by using a variational formulation. The boundary constraint on $\partial\Omega_{\rm sct}$ is defined as
\begin{equation}
\begin{aligned}
\mathcal L_{\rm bd}&=\int_{\mathbf r\in\partial\Omega_{\rm sct}}\left|\left.\frac{\partial u(\mathbf r)}{\partial \mathbf n(\mathbf r)}\right|_{r=R}\right|^2{\rm d}\mathbf r\\
&=\int_{\mathbf r\in\partial\Omega_{\rm sct}}\left|\left.\frac{\partial u_{\rm inc}(\mathbf r)}{\partial \mathbf n(\mathbf r)}\right|_{r=R}+\left.\frac{\partial u_{\rm sct}(\mathbf r)}{\partial \mathbf n(\mathbf r)}\right|_{r=R}\right|^2{\rm d}\mathbf r.\\
\end{aligned}
\label{eq:17}
\end{equation}
Consider the boundary constraint at the microphone positions with substituting (\ref{eq:7}) and (\ref{eq:15}) into (\ref{eq:17}), we obtain
\begin{equation}
\begin{aligned}
\mathcal L_{\rm bd}&\simeq\sum_{m=1}^M \left|\left.\frac{\partial u_{\rm inc}(\mathbf r_m)}{\partial \mathbf n(\mathbf r_m)}\right|_{r=R}+\left.\frac{\partial u_{\rm sct}(\mathbf r_m)}{\partial \mathbf n(\mathbf r_m)}\right|_{r=R}\right|^2\\
&=\left(\frac{\partial \mathbf K_{\rm inc}}{\partial \hat{\mathbf n}} \boldsymbol \alpha+\frac{\partial \mathbf K_{\rm sct}}{\partial \hat{\mathbf n}} \boldsymbol \beta\right)^\mathsf H\left(\frac{\partial \mathbf K_{\rm inc}}{\partial \hat{\mathbf n}} \boldsymbol \alpha+\frac{\partial \mathbf K_{\rm sct}}{\partial \hat{\mathbf n}} \boldsymbol \beta\right)\\
&=\boldsymbol \alpha^\mathsf H\frac{\partial \mathbf K_{\rm inc}^\mathsf H}{\partial \hat{\mathbf n}}\frac{\partial \mathbf K_{\rm inc}}{\partial \hat{\mathbf n}}\boldsymbol \alpha
+\boldsymbol \alpha^\mathsf H\frac{\partial \mathbf K_{\rm inc}^\mathsf H}{\partial \hat{\mathbf n}}\frac{\partial \mathbf K_{\rm sct}}{\partial \hat{\mathbf n}}\boldsymbol \beta\\
&\qquad\qquad+\boldsymbol \beta^\mathsf H\frac{\partial \mathbf K_{\rm sct}^\mathsf H}{\partial \hat{\mathbf n}}\frac{\partial \mathbf K_{\rm inc}}{\partial \hat{\mathbf n}}\boldsymbol \alpha
+\boldsymbol \beta^\mathsf H\frac{\partial \mathbf K_{\rm sct}^\mathsf H}{\partial \hat{\mathbf n}}\frac{\partial \mathbf K_{\rm sct}}{\partial \hat{\mathbf n}}\boldsymbol \beta,
\end{aligned}\\
\label{eq:18}
\end{equation}
where
\begin{equation}
\begin{aligned}
\left[\frac{\partial\mathbf K_{\rm inc}}{\partial \hat{\mathbf n}}\right]_{m,m'}=\frac{\partial\kappa_{\rm inc}(\mathbf r_m,\mathbf r_{m'})}{\partial\mathbf n(\mathbf r_m)}
\end{aligned}
\label{eq:19}
\end{equation}
and
\begin{equation}
\begin{aligned}
\left[\frac{\partial \mathbf K_{\rm sct}}{\partial\hat{\mathbf n}}\right]_{m,m'}&=
\frac{\partial\kappa_{\rm sct}(\mathbf r_m,\mathbf r_{m'})}{\partial\mathbf n(\mathbf r_m)}\\
&=\sum_{\nu=0}^{N_{\rm ext}}\sum_{\mu=-\nu}^\nu \psi'_{\nu,\mu}(\mathbf r_m)\overline{\psi_{\nu,\mu}(\mathbf r_{m'})}\xi_\nu(R).
\end{aligned}
\label{eq:20}
\end{equation}
Then, by solving 
\begin{equation}
\frac{\partial \mathcal L_{\rm bd}}{\partial \boldsymbol \beta^\mathsf H}=\underbrace{\frac{\partial \mathbf K_{\rm sct}^\mathsf H}{\partial \hat{\mathbf n}}\frac{\partial \mathbf K_{\rm inc}}{\partial \hat{\mathbf n}}}_{\mathbf M_{\rm mix}}\boldsymbol \alpha
+\underbrace{\frac{\partial \mathbf K_{\rm sct}^\mathsf H}{\partial \hat{\mathbf n}}\frac{\partial \mathbf K_{\rm sct}}{\partial \hat{\mathbf n}}}_{\mathbf M_{\rm sct}} \boldsymbol \beta=\mathbf 0,
\label{eq:21}
\end{equation}
we obtain the relationship between $\boldsymbol \alpha$ and $\boldsymbol \beta$ as
\begin{equation}
\boldsymbol \beta=-\mathbf M_{\rm sct}^{-1}\mathbf M_{\rm mix}\boldsymbol \alpha.
\label{eq:22}
\end{equation}
From (\ref{eq:16}), the optimization problem is defined as
\begin{equation}
\begin{aligned}
\underset{{\boldsymbol \alpha,\boldsymbol\beta}}{\rm minimize}\ \mathcal J(\boldsymbol \alpha,\boldsymbol\beta)\coloneqq&\|\mathbf s-\mathbf K_{\rm inc} \boldsymbol \alpha-\mathbf K_{\rm sct}\boldsymbol\beta\|^2\\
&\qquad+\lambda_1\boldsymbol \alpha^\mathsf H\mathbf K_{\rm inc}\boldsymbol \alpha+\lambda_2\boldsymbol \beta^\mathsf H\mathbf K_{\rm sct}\boldsymbol \beta.
\end{aligned}
\label{eq:23}
\end{equation}
Substituting (\ref{eq:22}) into (\ref{eq:23}), the optimization problem can be written as a function of $\boldsymbol \alpha$ as
\begin{equation}
\begin{aligned}
&\underset{{\boldsymbol \alpha}}{\rm minimize}\ \mathcal J(\boldsymbol \alpha)\coloneqq \|\mathbf s-\underbrace{(\mathbf K_{\rm inc}-\mathbf K_{\rm sct}\mathbf M_{\rm sct}^{-1}\mathbf M_{\rm mix})}_{\mathbf A}\boldsymbol \alpha\|^2\\
&\qquad+\boldsymbol \alpha^\mathsf H\underbrace{[\lambda_1\mathbf K_{\rm inc}+\lambda_2(\mathbf M_{\rm sct}^{-1}\mathbf M_{\rm mix})^\mathsf H\mathbf K_{\rm sct}\mathbf M_{\rm sct}^{-1}\mathbf M_{\rm mix}]}_{\mathbf Q}\boldsymbol \alpha,
\end{aligned}
\label{eq:24}
\end{equation}
and its solution is obtained as 
\begin{equation}
\hat{\boldsymbol\alpha}=(\mathbf A^\mathsf H\mathbf A+\mathbf Q)^{-1}\mathbf A^\mathsf H\mathbf s
\label{eq:25}
\end{equation}
by solving $\frac{\partial \mathcal J(\boldsymbol{\alpha})}{\partial \boldsymbol \alpha^\mathsf H}=\mathbf 0$.

\section{Evaluation}
\label{sec:5}
\subsection{Development of a RSMA}
\label{sec:5:1}
\begin{figure}[t]
\centering
\centerline{\includegraphics[width=0.45\columnwidth]{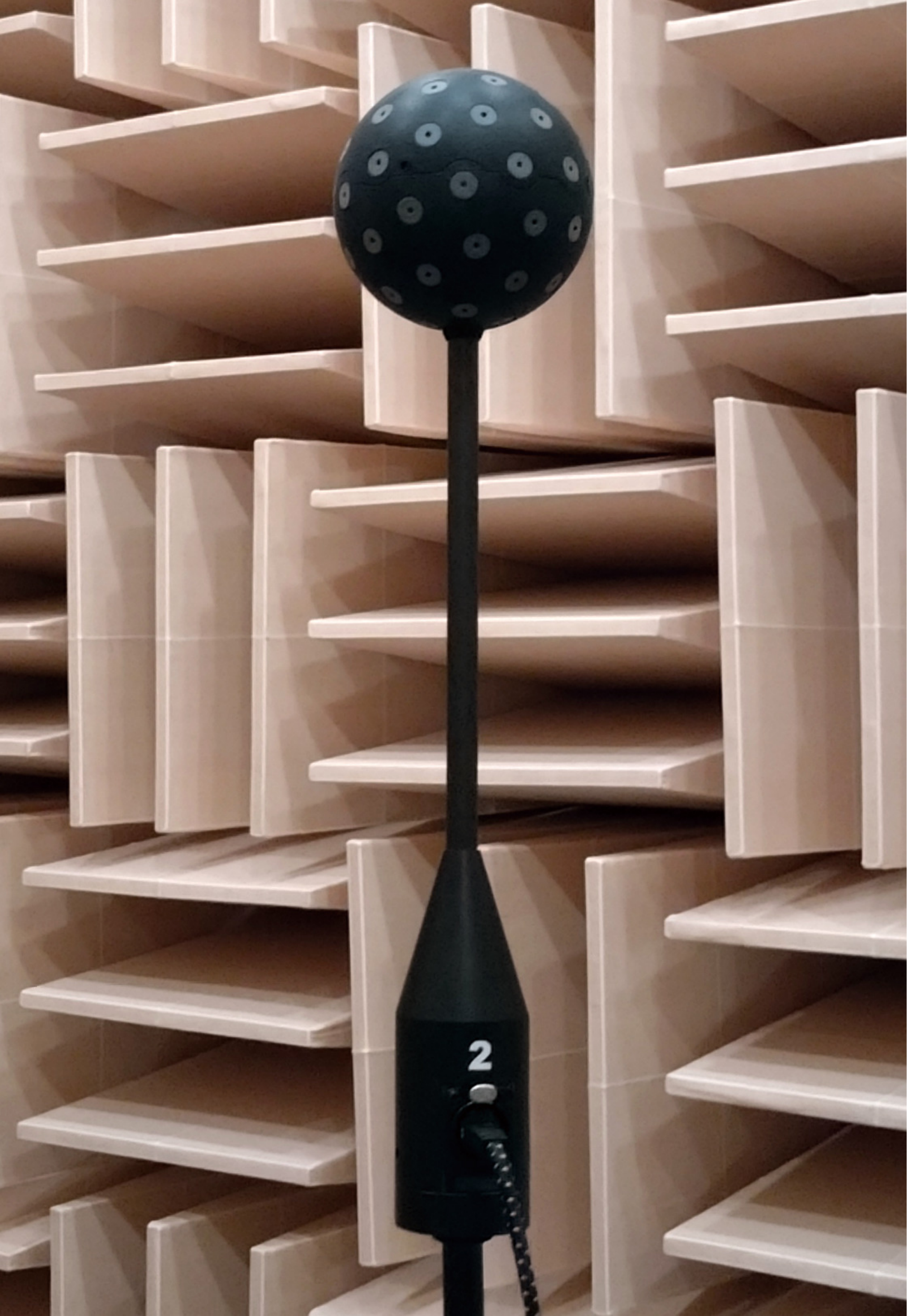}}
\caption{Our developed RSMA.}
\label{fig:1}
\end{figure}
We developed a RSMA for evaluating the methods. 60 Micro Electro Mechanical Systems (MEMS) \cite{mems} microphones are arranged on a rigid sphere with a radius of $0.05$ m based on a spherical $t$-design \cite{hardin1996mclaren}, ensuring complete orthogonality of spherical harmonics up to the 5th order. 
We employ Dante \cite{dante}, which is a proprietary audio-over-IP networking technology, and A${}^2$B \cite{a2b}, which is a digital audio bus system designed for automotive applications. The signals captured by the MEMS microphones are converted to Dante via an A${}^2$B interface, with the entire system operating in the digital domain. 
\Cref{fig:1} illustrates the physical configuration of the RSMA.
\subsection{Experimental setup}
\label{sec:5:2}
\begin{figure}[t]
\centering
\centerline{\includegraphics[width=0.9\columnwidth]{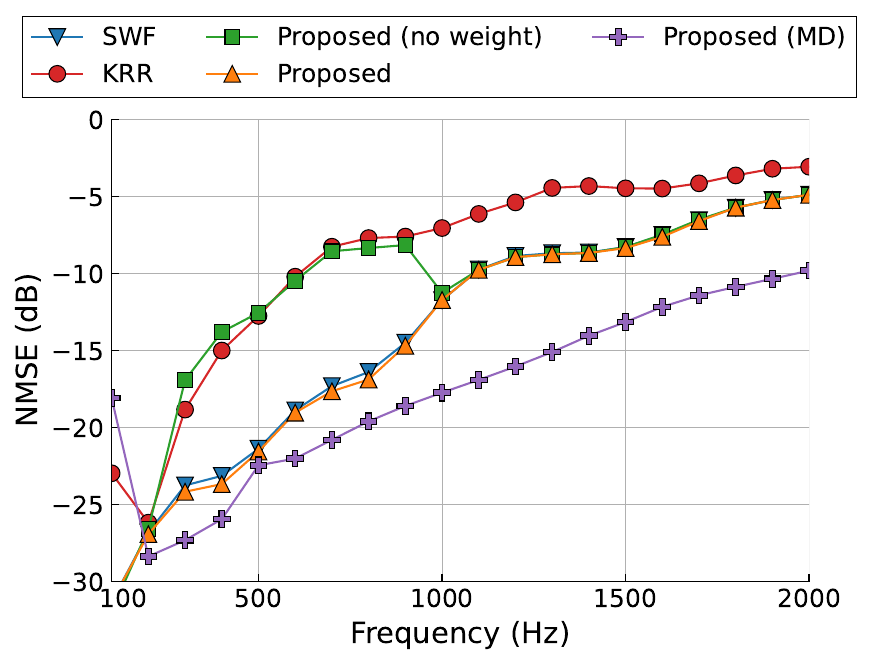}}
\caption{NMSE (dB) against frequency in the case of numerical simulation.}
\label{fig:2}
\end{figure}

\begin{figure}[t]
\centering
\centerline{\includegraphics[width=0.9\columnwidth]{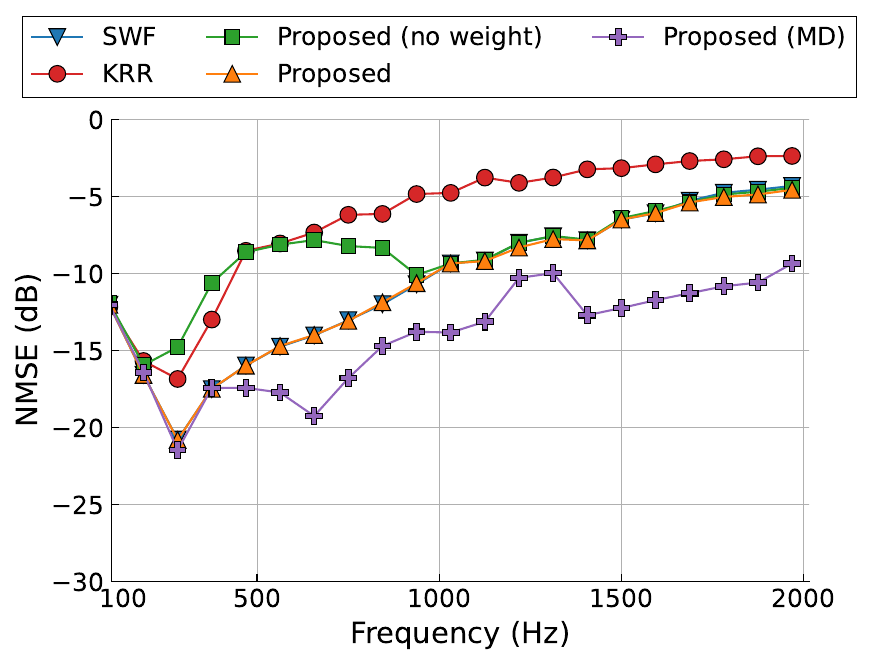}}
\caption{NMSE (dB) against frequency in the case of real-environment measurements.}
\label{fig:3}
\end{figure}
\begin{figure*}[t]
\centering
\centerline{\includegraphics[width=2\columnwidth]{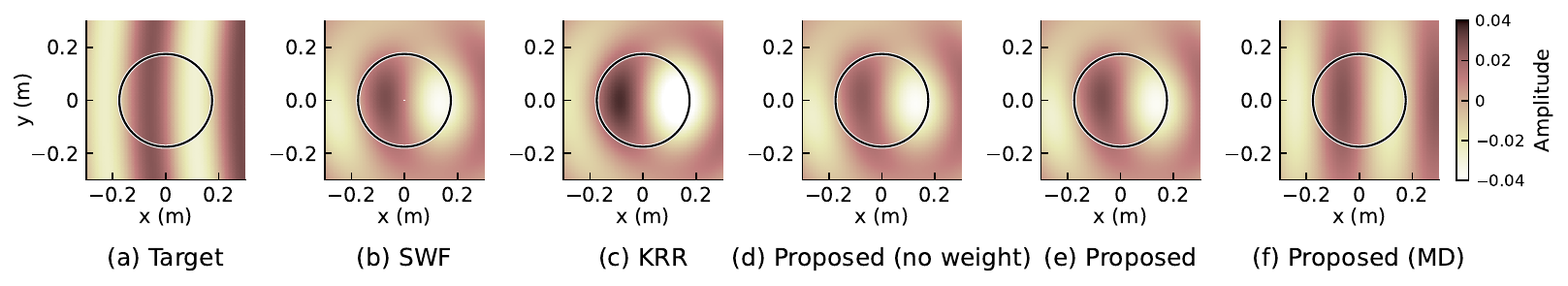}}
\caption{The real part of the estimated sound field on $xy$-plane in the case of numerical simulation at 1 kHz. The black circle indicates the target region.}
\label{fig:4}
\end{figure*}
\begin{figure*}[t]
\centering
\centerline{\includegraphics[width=2\columnwidth]{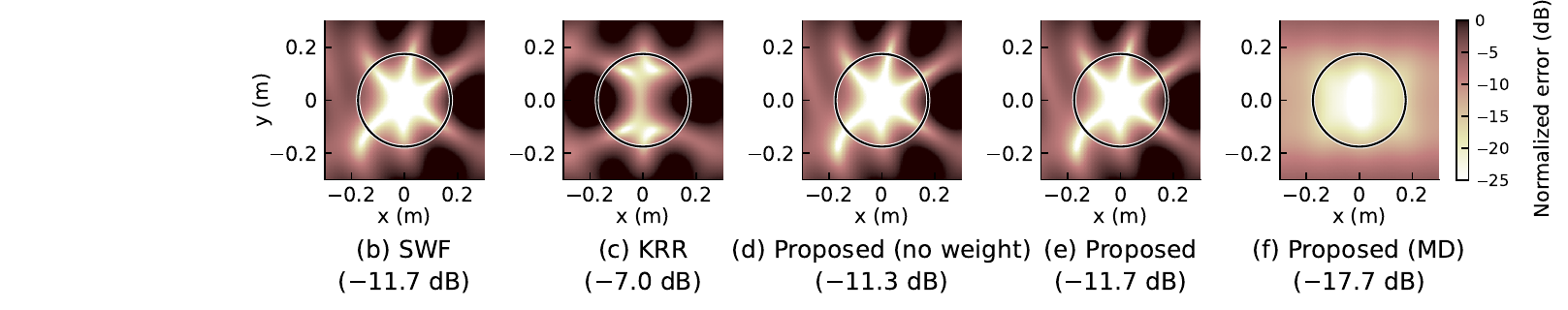}}
\caption{Normalized error distribution on $xy$-plane in the case of numerical simulation at 1 kHz. NMSEs are denoted below.}
\label{fig:5}
\end{figure*}
Numerical simulations and real-environment measurements were conducted to evaluate the proposed method. In both cases, the center of the RSMA was defined as the origin, and a single sound source was placed at the position $(3.0,0,0)\ {\rm m}$. The sound speed $c$ was set to $340.26$ m/s based on the measured room temperature of $14.6\ {}^\circ \rm C$ for both cases. In the numerical simulation, a free-field (anechoic) environment was assumed, and the sound pressure at each microphone position was calculated using (\ref{eq:4}) with $N=50$ assuming the amplitude of the sound source was equal to one. \Add{Additionally, Gaussian noise was added so that the SNR would be 20 dB.} The target region $\Omega$ was set to a spherical region with a radius of $0.175$ m, and $I=1000$ evaluation points were sampled uniformly within the sphere. The sound pressure at the evaluation points were calculated using the Green's function. For the real-environment case, the measurements were conducted in an anechoic chamber and the evaluation points were defined as a discretized grid within a rectangular volume of $0.35\ {\rm m} \times0.35\ {\rm m} \times0.20\ {\rm m}$, centered at the origin, with a spacing of $0.05$ m. We obtain the sound pressures at the RSMA and the evaluation points using the FFT of measured impulse responses. 

We compared the proposed method with the SWF expansion and the KRR without boundary conditions. The estimated sound fields for the SWF expansion were obtained using (\ref{eq:6}) and then (\ref{eq:2}) where $N$ was set to $5$. In the KRR without boundary conditions, the Bessel kernel was used, the sound field was estimated via (\ref{eq:9}) and then (\ref{eq:7}) where $N$ was set to $5$. Hereafter, they are denoted by {\bf SWF} and {\bf KRR}, respectively. As a proposed method, we compared three configurations: first, we used the Bessel kernel as the kernel function for the incident sound field and used \Add{(\ref{eq:14})} as $\xi_\nu(R)$; then, we here also used the Bessel kernel but $\xi_\nu(R)=1$; for the third configuration, we employed the MD kernel and $\xi_\nu(R)$ was set to \Add{(\ref{eq:14})}. In all cases, $N_{\rm ext}$ was set to 5. They are denoted by {\bf Proposed}, {\bf Proposed (no weight)}, and {\bf Proposed (MD)}, respectively. The regularization parameters for {\bf SWF}, {\bf KRR}, {\bf Proposed}, and {\bf Proposed (no weight)} were chosen from $10^l$ with $l\in\mathbb Z([-10,5])$ so that the NMSE was minimized for each frequency. 

In {\bf Proposed (MD)}, $\{\gamma_q\}_{q=1}^Q$ were optimized using the proximal gradient method assuming the spatial sparsity of the incident sound field as 
\begin{equation}
\gamma_q\leftarrow \mathcal S_{\eta_{\gamma}\tau}\left(\gamma_q-\eta_{\gamma}\frac{\partial \mathcal L_{\rm \Add{MD}}}{\partial\gamma_q}\right)
\label{eq:26}
\end{equation}
where $\mathcal S_{\eta_{\gamma}\tau}(\cdot)$ is a soft-thresholding operator defined as 
\begin{equation}
\mathcal S_{\eta\tau}(x)=\max(0,x-\eta\tau),
\label{eq:27}
\end{equation}
$\mathcal L_{\rm \Add{MD}}$ is \Add{an} objective function for update, $\eta_{\gamma}$ is the learning rate for $\{\gamma_q\}_{q=1}^Q$, and $\tau$ is the hyperparameter which decides the sparsity of $\{\gamma_q\}_{q=1}^Q$. In addition, $\{\zeta_q\}_{q=1}^Q$ were also updated using the gradient method with the learning rate of $\eta_{\zeta}$. On the other hand, $\{\mathbf d_q\}_{q=1}^Q$ were arranged by \Add{Lebedev} quadrature of order $7$ and fixed during the optimization. We experimentally set the hyperparameters as $\tau=10^{-2}$, $\eta_{\gamma}=10^{-1}$, $\eta_{\zeta}=10^2$ and employ leave-one-out cross validation loss as $\mathcal L_{\rm \Add{MD}}$. In order to ensure numerical stability during the parameter update, the regularization term in (\ref{eq:25}) was fixed by setting $\mathbf Q=\lambda \mathbf I$, where $\lambda=10^{-2}$. We set $\gamma_q=1/Q$ and $\zeta_q=20$ for all $q\in\{1,\dots,Q\}$ as the initial condition. The stopping criterion for the updates was set to the number of iterations, which was fixed at $400$.

As an evaluation metric, the normalized mean squared error (NMSE) was defined as
\begin{equation}
{\rm NMSE}\coloneqq \frac{\sum_{i=1}^I |\hat{u}_{\rm inc}(\mathbf r_i)-u_{\rm inc}(\mathbf r_i)|^2}{\sum_{i=1}^I |u_{\rm inc}(\mathbf r_i)|^2},
\label{eq:28}
\end{equation}
where $\mathbf r_i$ denotes the evaluation position.
\subsection{Results}
\label{sec:5:3}
\Cref{fig:2} shows the NMSE (dB) against frequency for numerical simulations. The comparison between the NMSEs of {\bf KRR} and {\bf Proposed} demonstrates that the incorporation of the boundary constraint contributes to improved estimation of the incident sound field for all analyzed frequencies. In addition, the comparison between {\bf Proposed} and {\bf Proposed (no weight)} reveals the effectiveness of the weight in the kernel function for the scattered field. Compared to the {\bf SWF} which uses the boundary condition, {\bf Proposed} exhibits slightly lower NMSEs in the frequency range below 1 kHz. Furthermore, {\bf Proposed (MD)} shows the lowest NMSE across the entire frequency range compared to the other methods, demonstrating the effectiveness of simultaneous optimization of the kernel function parameters within the proposed framework.

\Cref{fig:3} shows the NMSE (dB) for the experiments in the anechoic chamber. In the experimental results, although the NMSE increased in the low-frequency range for all methods, the overall trend was consistent with the simulation results, confirming the effectiveness of the proposed method. As an example, we plotted the estimated sound field and the normalized error distribution on $xy$-plane at 1 kHz in \cref{fig:4} and \cref{fig:5}, respectively. These results indicate that {\bf Proposed (MD)} achieves higher estimation accuracy than the other methods even outside the target region.

\section{Conclusion}
\label{sec:6}
In this paper, we proposed a KRR based sound field estimation method introducing a boundary constraint and a KRR representation of the scattered sound field. Numerical simulations and measurements taken in an anechoic chamber have shown that the proposed method achieves higher estimation accuracy, especially when using the MD kernel for the incident sound field. 
\clearpage
\bibliographystyle{IEEEtran}
\bibliography{refs25}







\end{document}